# Reshape of the Bunch-by-bunch BPM Signal to a Turn-by-turn Matrix during the Fast RF Frequency-sweeping Time in Booster

**Xi Yang, Charles M. Ankenbrandt, James Lackey, and Vic Scarpine**

*Fermi National Accelerator Laboratory*

**Box 500, Batavia IL 60510**

## Abstract

The bunch-by-bunch BPM signal array has to be reshaped into a turn-by-turn (TBT) matrix in order to make the TBT beam position information of a single bunch available for further data analysis, such as via the Fourier transform to obtain the synchrotron tune and betatron tune, etc. Since the RF frequency sweeps more than 10 MHz in the first 8 ms of a Booster cycle, the revolution time decreases more than 20%. The number of data points for each Booster turn (BT) should be varied in the same pace with the revolution time since the sampling rate of a digital oscilloscope is usually fixed. And this can be done by pre-determining the relationship between the RF frequency and the time in a cycle via the curve fit.

## Introduction

The BPM system in Booster can read the average of about 15 bunches due to the electronic limitation.[1] However, the BPM detector has a sufficient bandwidth of 3 GHz to resolve a single Booster bunch.[2] The BPM detector has four striplines in a single housing to allow measurements in both planes at one location. *A* and *B* represent signals from the two striplines in the plane of motion. In the horizontal plane, *A* and *B* are signals from the outside stripline and the inside stripline respectively; and in the vertical plane, they represent signals from the top stripline and the bottom stripline. The beam position is proportional to ($A$-$B$)/($A$+$B$). In our experiment, a hybrid was used to produce output signals of $A$-$B$ and $A$+$B$ from input signals of $A$ and $B$ in the horizontal long 18 detector. A Tektronix TDS 7104 digital oscilloscope with a 1-GHz bandwidth was used to record signals of $A$-$B$ and $A$+$B$. Signals of $A$-$B$ and $A$+$B$ have a bunch-by-bunch resolution since they didn't go through the BPM electronics, which has a bandwidth of about 4 MHz.

In order to reshape signal arrays of $A$-$B$ and $A$+$B$ to TBT matrices, the number of data points for each BT during the period when the data were taken should be precisely known, and they are the same for both $A$-$B$ and $A$+$B$ since cable lengths of signals $A$ and $B$ are well matched. Here, the 1$^{st}$ point of the $i^{th}$ turn in the TBT matrix is the same with the $m_i^{th}$ point in the signal array. Since the sampling rate of the scope is a constant of $f_s$=2500 MHz and the Booster harmonic number ($h$) is 84, once the RF frequency at the $i^{th}$ turn ($f_{rf}(m_i)$) is known, the revolution period is known from '$h/f_{rf}(m_i)$', the number of data points in the $i^{th}$ turn ($N(m_i)$) can be calculated using eq.1.

$$N(m_i) = \left( h \big/ f_{rf}(m_i) \right) \cdot f_s. \tag{1}$$

Since the proton beam is accelerated from 400 MeV to 8 GeV in Booster in a time of 33.3 ms, while the RF frequency sweeps from 37.8 MHz to 52.9 MHz. The RF frequency increases more than 2 MHz during 5 ms to 6.5 ms after the injection. The 2$^{nd}$ – order polynomial fit was used to find the relationship between $f_{rf}(m_i)$ and $m_i$, as shown by eq.2, since the relationship between the RF frequency and the time in a Booster cycle is nonlinear.

$$f_{rf}(m_i) = A + B_1 \cdot m_i + B_2 \cdot m_i^2, \tag{2}$$

Also, eq.3 is used to calculate $m_i$.

$$\begin{aligned} m_i &= m_{i-1} + N_{i-1}(m_{i-1}), \text{ when } i > 1; \\ m_i &= 1, \qquad\qquad\qquad \text{ when } i = 1. \end{aligned} \tag{3}$$

## Experimental Result and Analysis

Signals $A$-$B$ and $A$+$B$ were recorded by a Tektronix TDS 7104 digital oscilloscope during the time of 5 ms to 6.6 ms in a MiniBooNE event at the extracted beam intensity of $4.0\times10^{12}$ protons. Since the beam notch (BN) was created at about 4 ms after the injection well before the data were taken for the purpose of reducing the extraction beam loss, it could be conveniently used as the indicator of a single BT in the data analysis, as shown in Fig.1. Eight BTs, which are evenly distributed in time along the signal array of *A*-*B*, were chosen and their RF frequencies were calculated using eq.1. The result is shown as the black curve in Fig.2. Afterwards, the relationship between the RF frequency $f_{rf}(m_i)$ and the point number $m_i$ in the signal array was determined by the curve fit using eq.2, as shown by the red curve in Fig.2, and the results are $A$= 42.34987, $B_1$= 7.0511E-7, and $B_2$= -5.09911E-15.

The TBT matrix is formed in such a way that each row corresponds to a single BT and the row number represents the turn number. Since the revolution time for a single BT got shorter during the time when the data were taken, the corresponding number of data points for a row also got smaller. The number of data points for the 1st turn is used to define the 2nd dimension of the TBT matrix, and the rest part of a row is filled by zero when the row number is greater than one. It is important for us to have a final adjustment for parameters $A$, $B_1$ and $B_2$ for the purpose of making each row of the TBT matrix have the same starting point relative to the circulating beam.

As an example, a constant of 4956 was used as the number of data points for each row and also the 2nd dimension of the matrix to reshape the signal array *A*-*B*, and the result is shown as the mountain range (MR) plot, Fig.3(a). The BN slews within tens of BTs and is indicated by the curved arrow. Finally, parameters $A$=42.34987, $B_1$= 7.0811E-7 and $B_2$=-7.1E-15 were used for reshaping signal arrays of *A*-*B* and *A*+*B* to TBT matrices. The result of the TBT, which is obtained from the signal array *A*-*B*, is shown as the MR plot, Fig.3(b), and each beam bunch is clearly identifiable.

The bunch position is obtained using eq.4.

$$x_{i,j} = \frac{\max\left((A-B)_{i,j}\right) - \min\left((A-B)_{i,j}\right)}{\max\left((A+B)_{i,j}\right) - \min\left((A+B)_{i,j}\right)}. \tag{4}$$

Here, $i$ is the turn number, and $j$ is the bunch number in each row of the TBT matrix.

The 3th bunch from the turn number of $i$=251 to $i$=506, which is indicated by the double arrow in Fig.4(a), is used to perform the fast Fourier transform (FFT), and its unnormalized power spectrum is shown in Fig.4(b). The synchrotron frequency ($f_{syn}$) and the RF frequency ($f_{rf}$) around 5 ms are about 27 kHz and 42.7 MHz.[3] The synchrotron tune ($Q_s$) is estimated using eq.5, and the result is 0.053, which is consistent with the peak #1 in Fig.4(b).

$$Q_s = {f_{syn}} \Big/ \left( {f_{rf}} \big/ {h} \right). \tag{5}$$

If the peak #2 in Fig.4(b) is indeed caused by the betatron motion, the separation between the peak #2 and the peak #3 is equal to the synchrotron tune, which is consistent with the chromaticity-sideband prediction.[4] Whether or not the peak #2 is caused by the betatron motion needs to be further confirmed. For example, one can differentiate this by watching the movement of the peak #2 while changing the betatron tune via a different quad setup.

**Comment**

The synchrotron tune and betatron tune are clearly identifiable in the unnormalized power spectrum of a single-bunch TBT BPM data even when the data were taken at the normal operational condition. One expects that the synchrotron tune, betatron tune, and chromaticity sidebands would be more pronounced in the condition when circulating bunches are excited both transversely and longitudinally.

**Acknowledgement**

Authors give special thanks to Bill Pellico for providing raw BPM signals from the long-18 pickup.

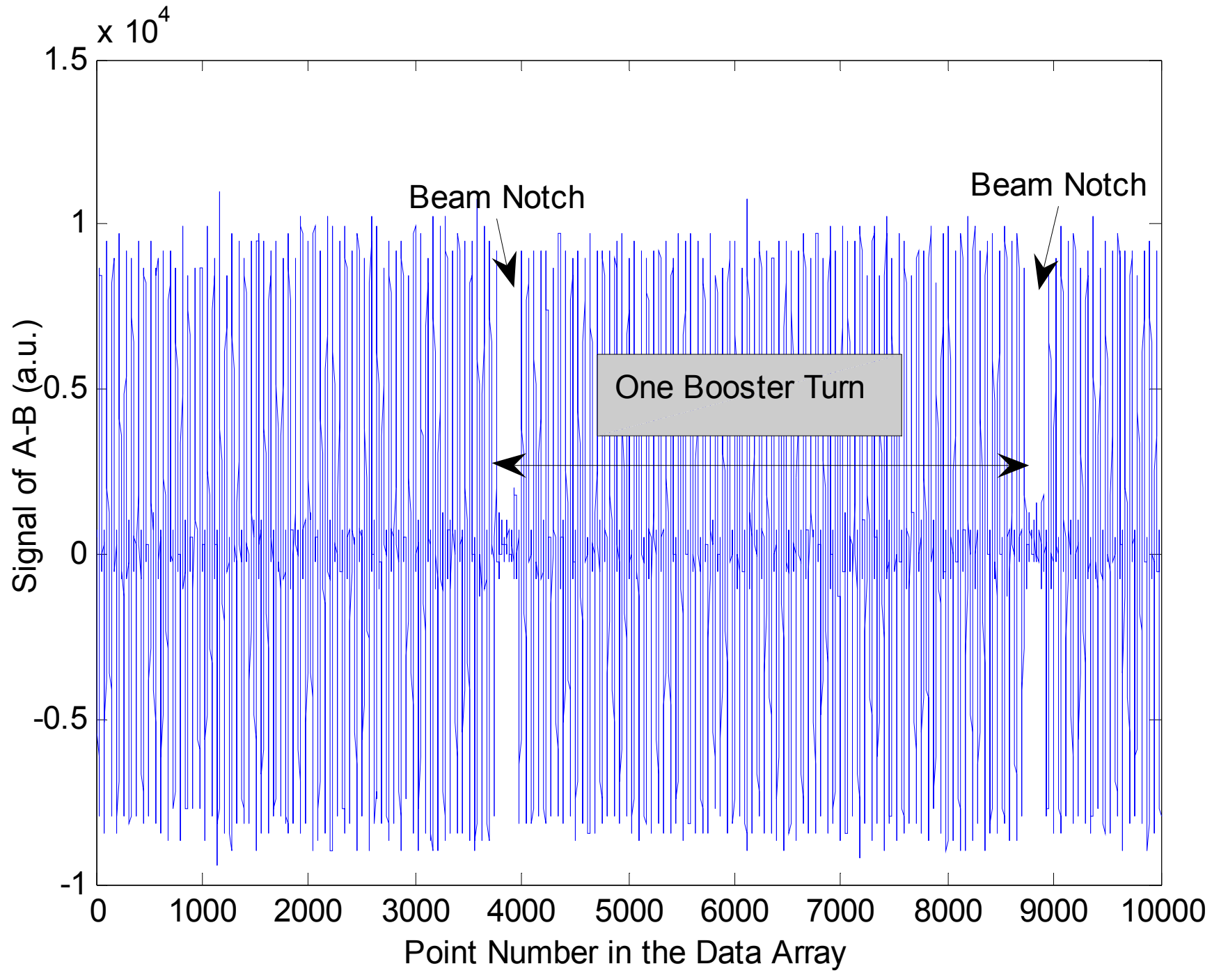

Fig. 1

Fig. 1 the signal of *A*-*B* in the first 10000 points.

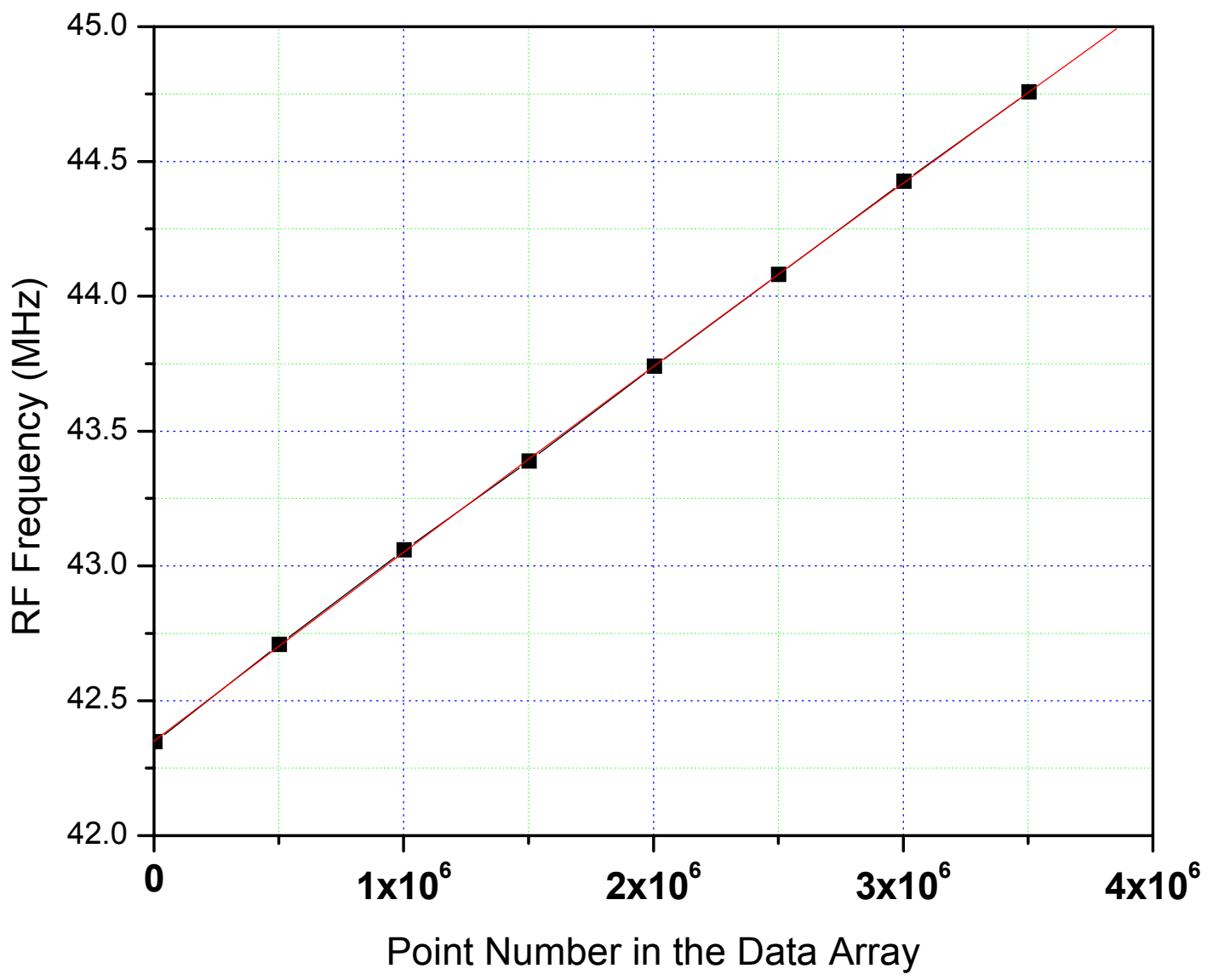

Fig.2

Fig. 2 RF frequency *vs*. the point number in the scope signals of *A*-*B* and *A*+*B*. The black curve and the red curve represent the measurement and the curve-fit result respectively.

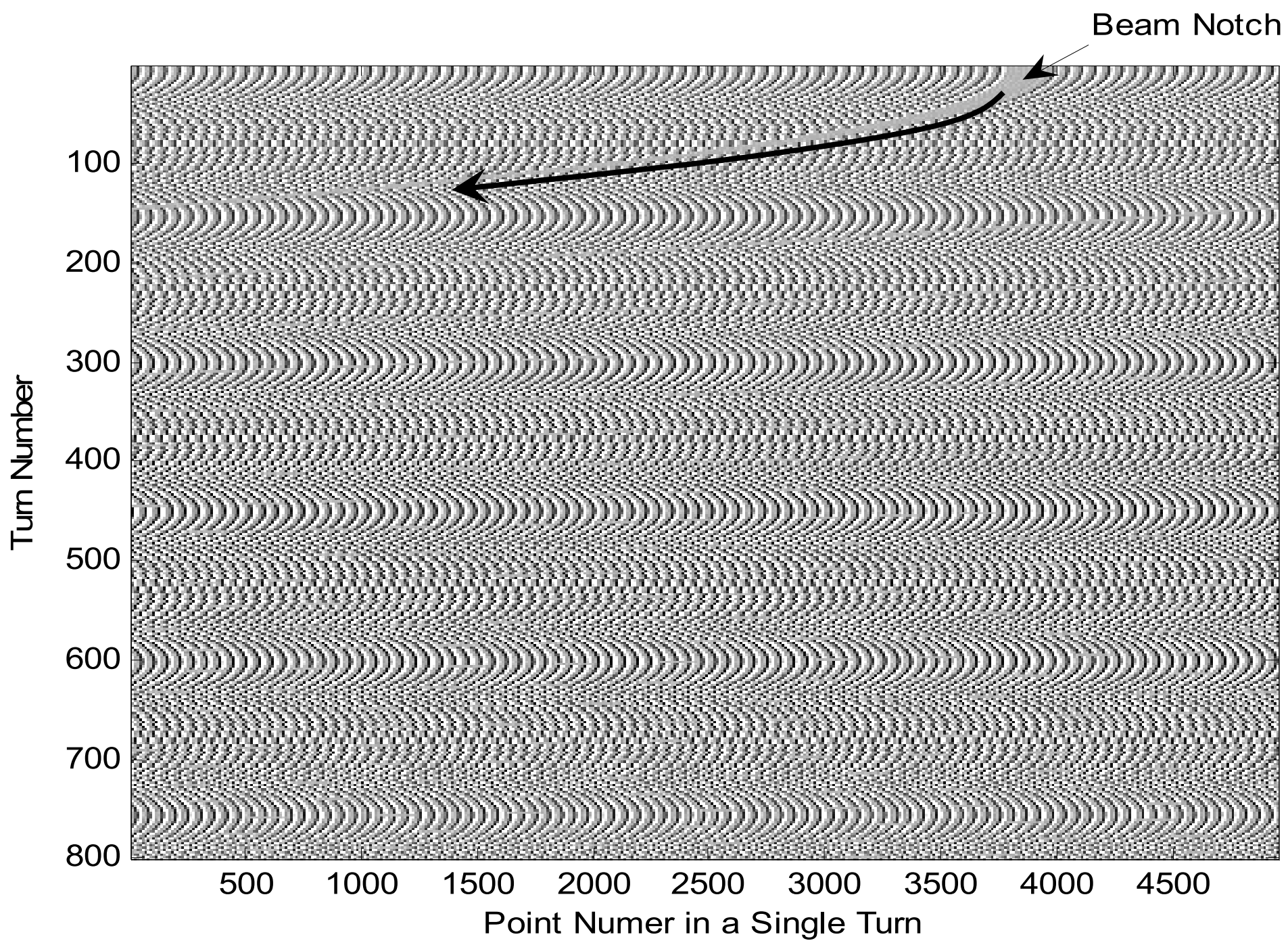

Beam Notch
Turn Number
100
200
300
400
500
600
700
800
500 1000 1500 2000 2500 3000 3500 4000 4500
Point Numer in a Single Turn

Fig. 3(a)

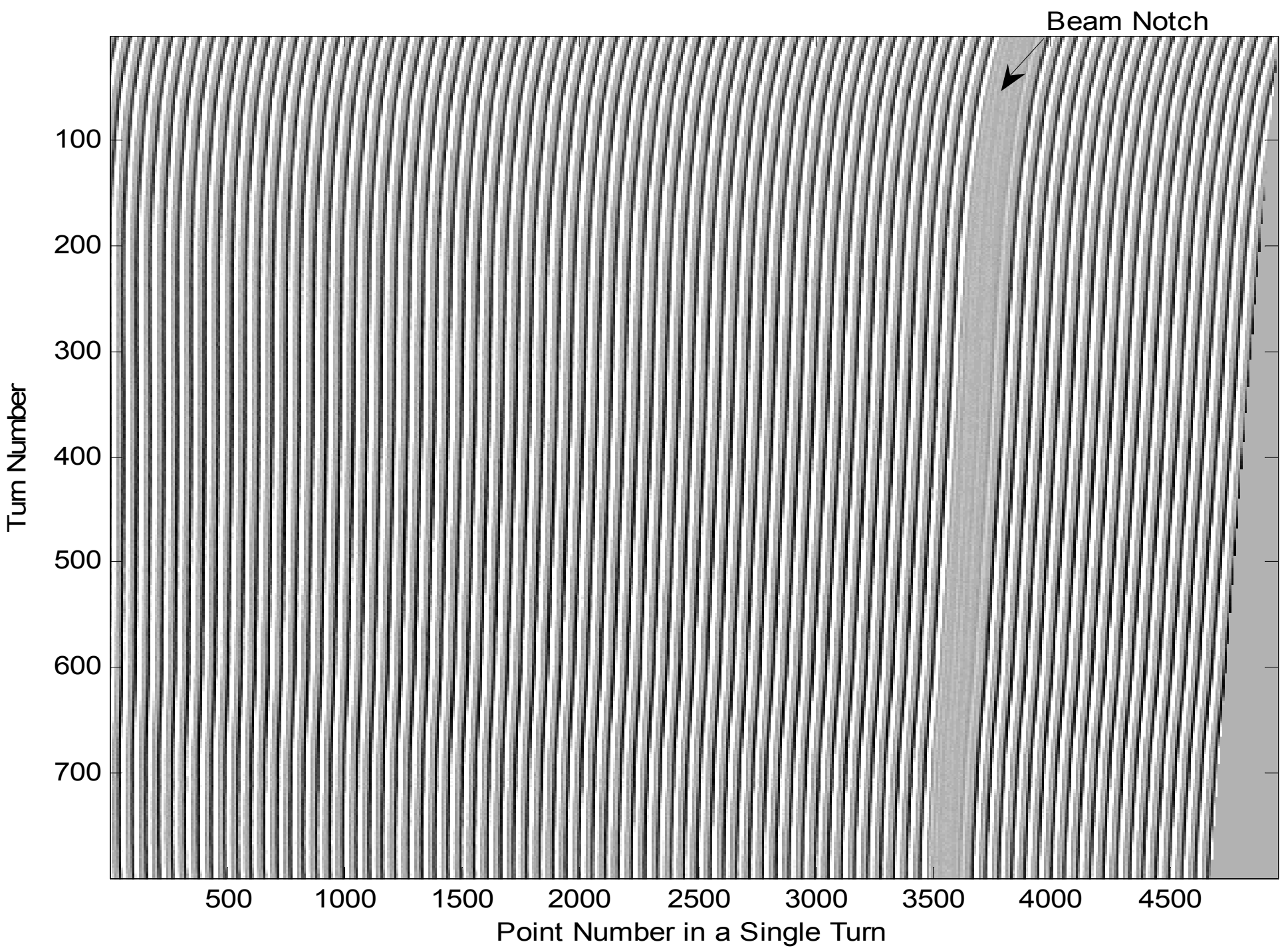

Beam Notch
Turn Number
100
200
300
400
500
600
700
500 1000 1500 2000 2500 3000 3500 4000 4500
Point Number in a Single Turn

Fig. 3(b)

Fig. 3(a) the MR plot of the TBT matrix, which was obtained from the signal of *A*-*B* when a constant of 4956 was used as the number of data points for each row and the 2nd dimension of the matrix.

Fig. 3(b) the MR plot of the TBT matrix, which was obtained from the signal of *A*-*B* when the number of signal points in each row was determined by the curve-fit result.

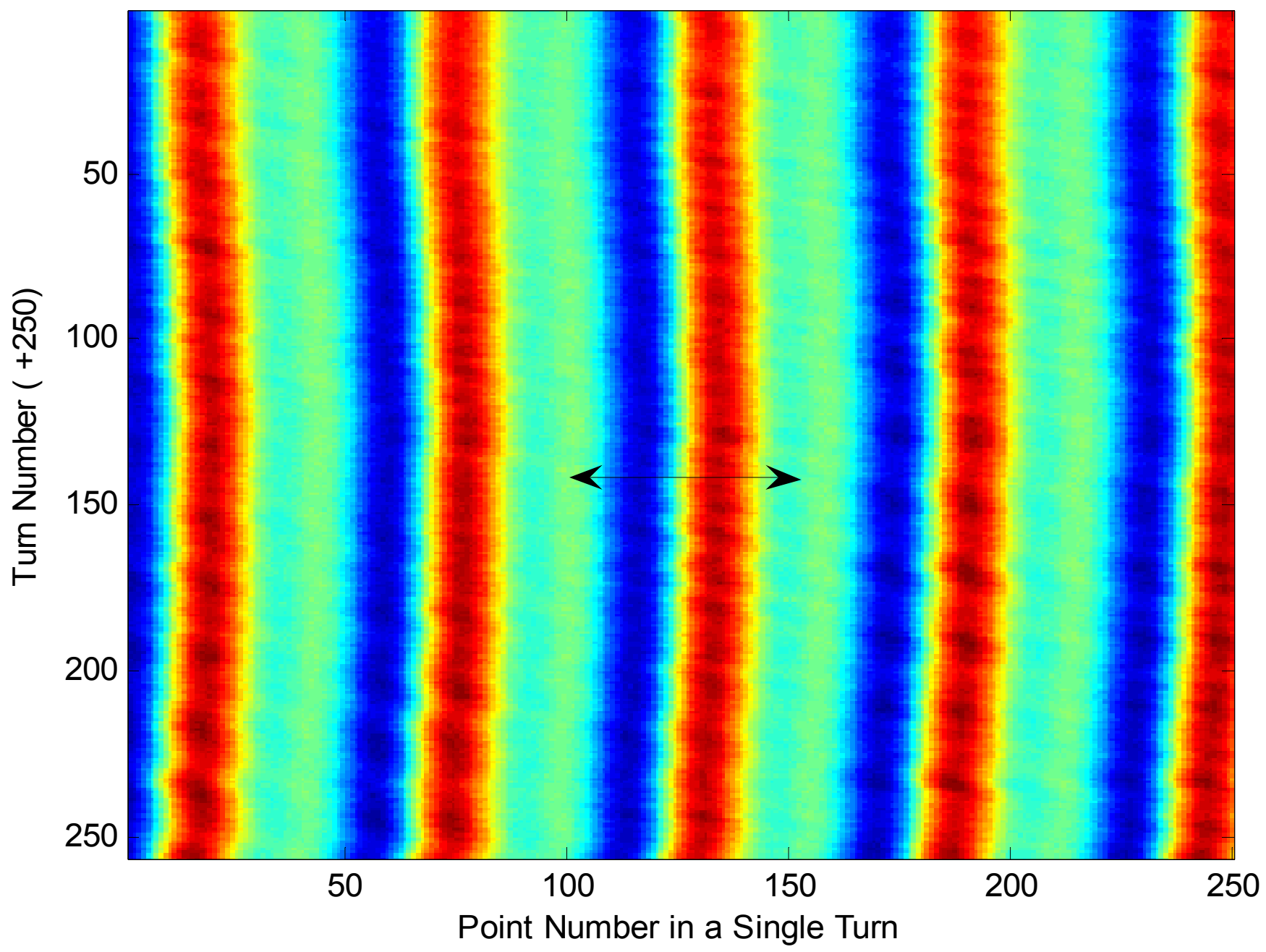

Fig. 4(a)

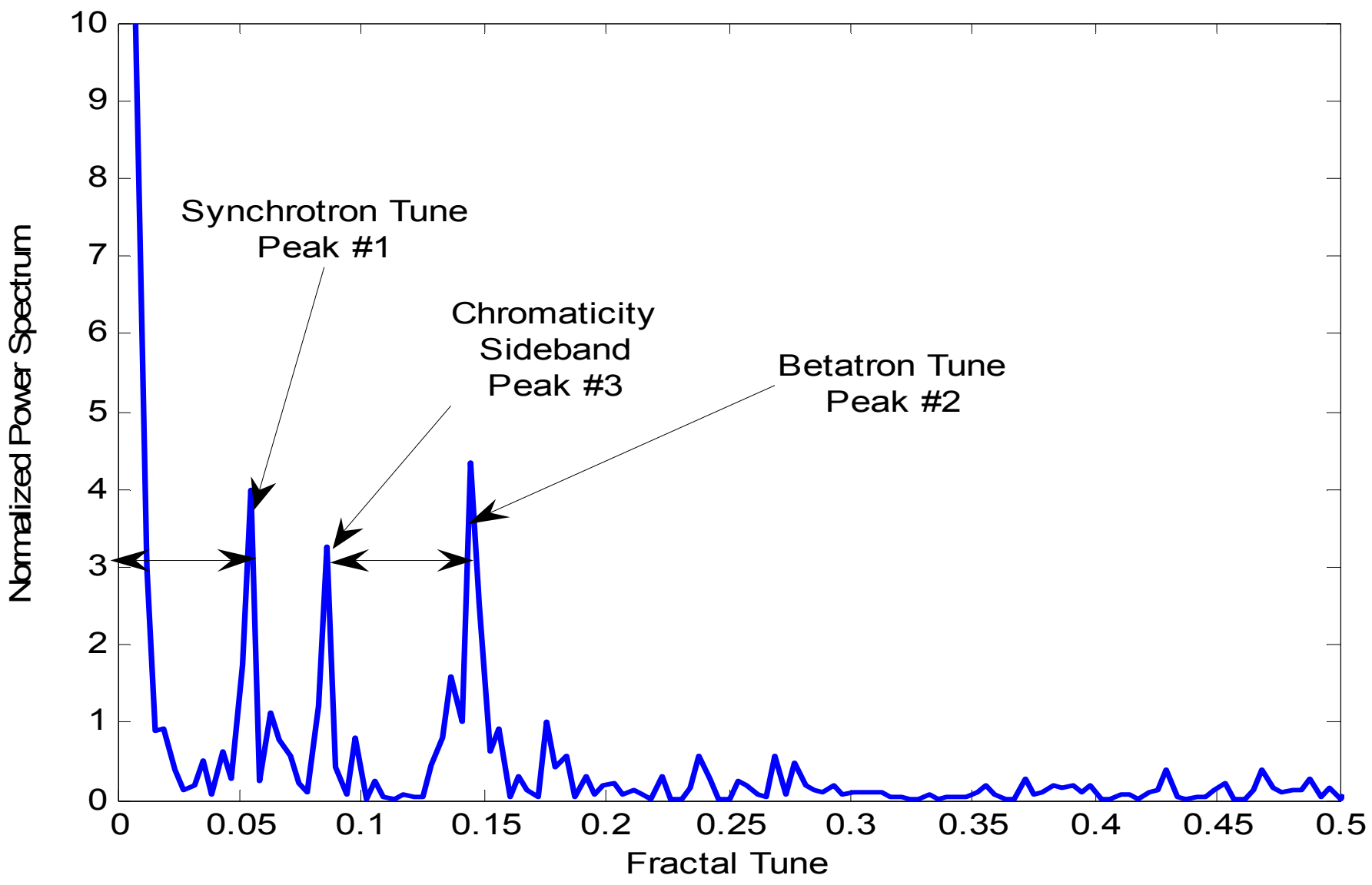

Fig. 4(b)

Fig. 4(a) the MR plot of the first five bunches from the turn number $i$=251 to $i$=506.

Fig. 4(b) the unnormalized power spectrum of the 3$^{th}$ bunch from the turn number $i$=251 to $i$=506.